\documentclass[%
reprint,
prb,
]{revtex4-2}

\usepackage{amsmath,amssymb}
\usepackage{graphicx}

\usepackage[caption=false]{subfig}

\usepackage[pdftex, hidelinks,
pdfauthor={Johannes Mellak},
pdftitle={no title},
pdfproducer={Latex with hyperref, or other system},
pdfcreator={pdflatex, or other tool}]{hyperref}

\usepackage{bm}  
\newcommand{\ve}[1]{{\bm{#1}}} 
\newcommand{\rh}{\hat{\rho}}
\newcommand{\si}{{\bm{\sigma}}}
\newcommand{\Li}{{\mathcal{L}}}
\newcommand{\vtheta}{{\bm{\theta}}}
\newcommand{\statexp}[1]{{\langle\langle #1 \rangle\rangle}}
\newcommand{\ket}[1]{ |  #1 \rangle}
\newcommand{\bra}[1]{ \langle  #1 |}

\usepackage{bbold}
\usepackage{color}

\usepackage{cancel}
\usepackage{ulem}

\begin{document}

\title{Quantum Transport in Open Spin Chains using Neural-Network Quantum States} 

\author{Johannes Mellak}
\email{mellak@tugraz.at}
\author{Enrico Arrigoni}
\author{Thomas Pock}
\altaffiliation{Institute of Computer Graphics and Vision, Graz University of Technology, Inffeldgasse 16/II A-8010 Graz}
\author{Wolfgang von der Linden}
\affiliation{Institute of Theoretical and Computational Physics, Graz University of Technology, Petersgasse 16/II
	A-8010 Graz}

\date{\today}

\begin{abstract}

In this work we study the treatment of asymmetric open quantum systems with neural-networks based on the restricted Boltzmann machine. 
In particular, we are interested in the non-equilibrium steady state current in the boundary-driven (anisotropic) Heisenberg spin chain. 
We address previously published difficulties in treating asymmetric dissipative systems with neural-network quantum states and Monte-Carlo sampling and present an optimization method and a sampling technique that 
can be used to obtain high-fidelity steady state approximations of such systems. 
We point out some inherent symmetries of the Lindblad operator under consideration and exploit them during sampling. 
We show that local observables are not always a good indicator of the quality of the approximation and finally present results for the spin current that are in agreement with known results of simple open Heisenberg chains.

\end{abstract}

\maketitle

\section{Introduction}

Due to the increasing importance of nanoscale 
devices 
the theoretical  treatment of open quantum systems, coupled to an environment or  \emph{bath}, have become of great interest.
In many cases, these quantum scale 
devices exhibit pronounced correlation effects. Strongly correlated quantum many-body systems, however, 
have an exponentially growing Hilbert space comprising all possible many-body configurations and the applicability of exact numerical methods, in particular for non-equilibrium situations, are strongly limited.

In recent years so-called neural-network quantum states (NQS), using artificial neural networks as variational wavefunctions, were shown to be able to efficiently encode correlations in various strongly correlated quantum many-body systems in equlibrium \cite{carleo_solving_2017, pfau_ab-initio_2020, hermann_deep-neural-network_2020, choo_study_2019, luo_backflow_2019, nomura_restricted_2017}. 
This scheme has been extended to describe the non-equilibrium steady state  \cite{vicentini_variational_2019, nagy_variational_2019, yoshioka_constructing_2019} 
and the dynamics \cite{hartmann_neural-network_2019} of open quantum systems
based on Restricted Boltzmann Machines (RBM) or using a probabilistic formulation together with direct sampling of autoregressive neural networks \cite{luo_autoregressive_2021}. 
The optimization methods to obtain the neural networks parameters are based on Monte-Carlo (MC) sampling and can therefore be easily parallelized, and share many similarities with general machine learning (ML) problems that often use stochastic approximations of gradients.

In a previous work \cite{kaestle_sampling_2021} it has been shown that asymmetric (inhomogeneous) open quantum systems 
are quite hard to simulate by machine learning algorithms based on conventional Metropolis sampling. 
The authors introduced a hybrid sampling strategy which takes into account the inhomogeneity of the system. 
This resulted in a better accuracy  
of local quantities such as the magnetisation. 
However, asymmetrically driven systems support a steady state current (in case of Ref.~\cite{kaestle_sampling_2021} a spin current), which is often the most important quantity one is interested in. 

In this paper we show that an accurate value of the current is much harder to obtain by conventional machine learning approaches addressed so far, including the one in Ref.~\cite{kaestle_sampling_2021}. In this sense, we illustrate that accurate values for local quantities, such as the magnetization,  do not necessarily imply an accurate (high-fidelity) steady-state density matrix or an accurate (spin) current. 
We therefore propose a targeted 
optimization scheme that is capable, in case of the RBM state, to obtain high-fidelity steady states for asymmetric systems. 
Moreover, we introduce a sampling procedure that yields an 
unbiased stochastic 
approximation 
of the gradients of the cost function, which allows to obtain an accurate steady state.
This enables us to achieve accurate results for the spin current in the model systems under consideration.

\section{Neural Network Representation of the Many-Body Density Operator}

We now briefly describe the approach of the pioneering 
publications \cite{torlai_latent_2018, vicentini_variational_2019, nagy_variational_2019, hartmann_neural-network_2019} to treat the computation of the steady-state density matrix as an optimization problem using neural networks. 
We consider a finite system, described by a hamiltonian $H$, coupled to a Markovian environment. The time evolution of the reduced density operator 
is described  by a master equation in Lindblad form
\begin{equation}
	\frac{d\rh}{dt} = \Li \rh 
	= -i \left[H, \rh \right] 
	+ \sum_k \big( L_k \rh L_k^\dagger - \frac{1}{2} \{ L_k^\dagger L_k, \rh \} \big)
	\label{eq:lindblad}
\end{equation}
with the 
\emph{jump} 
operators $L_k$ 
describing dissipation 
and the driving mechanism 
responsible for a non-unitary time evolution.
The non-equilibrium steady-state (NESS) 
follows from 
$d\rh / dt = 0$ and corresponds to the eigenoperator
of the super-operator $\Li$ with eigenvalue 0. 
In Reference \cite{weimer_variational_2015} a variational scheme 
was introduced to obtain the steady-state by minimizing the 
trace norm of the time derivative in Eq.~(\ref{eq:lindblad}).

In \cite{vicentini_variational_2019, hartmann_neural-network_2019, nagy_variational_2019} this approach was applied to a RBM ansatz for the density operator
$ \rh_{\vtheta} = \sum_{\si \si'} \rho_{\vtheta}(\si, \si') |\si\rangle\langle\si'|$, and
the minimization was achieved by varying the variational parameters $\vtheta$. 
Here, $\ket{\si}$ is a complete orthonormal many-body basis for the  hamiltonian $H$ under consideration. 
%
%
As described in Reference \cite{vicentini_variational_2019},
we can also use the expectation value of $\Li^\dagger \Li $  
as the cost function to minimize
\begin{align}
C(\vtheta) &= 
\frac
	{\text{Tr} \left[\rh_{\vtheta}^\dagger \Li^\dagger \Li \rh_{\vtheta} \right]}
	{\text{Tr} \left[\rh_{\vtheta}^\dagger \rh_{\vtheta} \right]}
= \frac
	{|| \Li \rh_{\vtheta}||^2_2}
	{||\rh_{\vtheta}||^2_2}\\
&=  
\frac
	{\sum_{\si\si'}|\sum_{\tilde{\si}\tilde{\si}'} 
		\Li_{\si\si'\tilde{\si}\tilde{\si}' }
		\rho_\theta(\tilde{\si}, \tilde{\si}')|^2}
	{\sum_{\si\si'}| \rho_\theta(\si, \si')|^2} \label{eq:sum:HS} 
	\; .
\end{align}
The inner sum in equation \eqref{eq:sum:HS} can be carried out exactly  because 
 the Lindblad matrix $\Li_{\si\si'\tilde{\si}\tilde{\si}' }$ is sparse. The outer sum, however, runs over the entire Hilbert space. To reduce this elaborate  part of the computation, 
the cost function  is rewritten  as an expectation value in terms of the 
joint probability 
\begin{align}\label{eq:}
p_\vtheta (\si,\si')  = \frac{|\rho_\theta(\si, \si')|^2}{\sum_{\bar{\si}\bar{\si}'}| \rho_\theta(\bar{\si}, \bar{\si}')|^2}\;,
\end{align}
resulting in 
\begin{equation}
\label{eq:cost}
\begin{split}
C(\vtheta) 
&= \sum_{\si\si'} p_\vtheta (\si,\si') 
	\left| \sum_{\tilde{\si}\tilde{\si}'}	\Li_{\si\si'\tilde{\si}\tilde{\si}' }
	\frac{\rho_\theta(\tilde{\si}, \tilde{\si}')}
	{\rho_\theta(\si, \si')}\right|^2 
	\; .
\end{split}
\end{equation}
The expectation value can now be estimated by a Monte-Carlo sample mean.

To solve the resulting optimization problem 
\begin{align}
\underset{\vtheta}{\text{argmin}} \, C(\vtheta)
\end{align}
the parameters are iteratively updated {using} the gradient $\nabla_\vtheta C$, which can {be expressed} in terms of 
$D_k = \frac{\partial}{\partial \theta_k} \text{ln} \rho_\vtheta$
and evaluated using the same Monte-Carlo samples (see Refs.~\cite{vicentini_machine_2021, nagy_variational_2019, hartmann_neural-network_2019}). 
Within the stochastic reconfiguration (SR) method \cite{sorella_weak_2007}, the parameters at iteration $t$ are updated according to 
\begin{equation}
	\label{eq:sgd and sr}
\vtheta_{t+1} = \vtheta_t - \eta \; S^{-1} \nabla_{\vtheta_t} C
\end{equation}
where $\eta$ is the step size. %
Here, $S$ stands for the covariance matrix 
\begin{equation}
S_{kk'} = \statexp{D_k^* D_{k'}} - \statexp{D_k^*} \statexp{D_{k'}} \;,
\end{equation}
where $\statexp{.}$ denotes the expectation value in terms of the probability density~$p_\vtheta(\si,\si')$.
If we replace the covariance  by  the identity matrix   the stochastic gradient descent (SGD) update step is recovered. 
We will use both of these as reference methods. 

Once the parameters of the neural network are determined, the expectation value of a physical observable $\hat{O}$ can be computed as follows. We first rewrite it as an expectation value in terms of the probability $p(\si) = \rho(\si, \si)$ 
\begin{equation}
	\label{eq:observable}
	\langle \hat{O} \rangle = \text{Tr} \; \{\hat{O} \rho \}
	= \sum_\si p(\si) \sum_{\si'} \frac{\langle \si | \hat{O} | \si'\rangle \rho(\si, \si')}{\rho(\si, \si)} \; .
\end{equation}
The expectation value is then estimated by a Monte-Carlo sample  drawn from $p(\si)$.
Again the inner sum typically only contains a small number of terms and can be carried out exactly. 

The idea of addressing many-body problems via neural networks was introduced in Ref.~\cite{carleo_solving_2017}, 
exploiting their expressive power to represent quantum correlations in many-body states. 
The goal is to efficiently represent a complex-valued wavefunction amplitude for a many-body configuration, described by the basis $\ket{\si}$ for the $N$ physical degrees of freedom (f.e. spins), which are taken as the network's input. 
For concreteness, we consider the spin-1/2 Heisenberg model and the basis $\ket{\si}$ is the eigenbasis of the $S^{z}$-part of the model.

A typical RBM consists of a visible layer $\ve{\sigma}$ and a hidden layer $\ve{h}$ of binary units with connections only between nodes belonging to different layers. 
The many-body wavefunction ansatz is written in terms of a Boltzmann-like distribution over a fictitious Ising-type energy $E(\ve{\sigma}, \ve{h}; \ve{\theta})$ whose interaction  coefficients $\ve{\theta}$ have to be optimized as the parameters of the network. 
The $M$ hidden units $\ve{h}$ can be traced out analytically \cite{carleo_solving_2017} leading to a variational ansatz for the wavefunction $\psi$ in the form 
\begin{equation}
\psi_\ve{\theta}(\ve{\sigma}) = \sum_{\ve{h}} \text{exp}(-E(\ve{\sigma}, \ve{h}; \ve{\theta})) \; . 
\end{equation}
Typically, it is expected that the ansatz improves with increasing hidden node density $\alpha = M / N$ because with increasing number of parameters arbitrary functions can be approximated \cite{carleo_solving_2017}. 

The neural-network density operator (NDO) that we use as the variational ansatz 
is obtained by tracing out the environment degrees of freedom in an extended system  \cite{torlai_latent_2018}. 
A \emph{purifying} wavefunction $\psi$ is introduced to describe
the state of the physical system extended by a fictitious bath with the ancillary quantum numbers $\ve{a}$. 
From the density matrix for a pure state in this extended system
the reduced density in the physical system is obtained by marginalizing over the bath degrees of freedom $\ve{a}$
\begin{equation}
	\label{eq:rho_ansatz}
	\rho(\si, \si') = \sum_{\ve{a}} \psi^*(\si, \ve{a}) \psi(\si', \ve{a}) 
	\; .
\end{equation}

For the extended variational wavefunction $ \psi(\si, \ve{a})$ we
follow the procedure of Ref.~\cite{torlai_latent_2018} of using two real-valued RBM $P_{\ve{\lambda}/\ve{\mu}}$, one for the real and one for the imaginary part, with their respective parameter sets $\ve{\lambda}$ and $\ve{\mu}$
\begin{equation}
	\label{eq:psi_split_real_imag}
	\psi_{\ve{\lambda}\ve{\mu}}(\si, \ve{a})  = \sqrt{P_\ve{\lambda}(\si, \ve{a})} \;  e^{i \Phi_\ve{\mu}(\si, \ve{a})}
\end{equation}
with $\Phi_\ve{\mu}(\si, \ve{a}) = \text{log}P_\ve{\mu}(\si, \ve{a}) /2$. 
The $K$ ancillary nodes $\ve{a}$, introduced to represent the bath, are taken as additional hidden units with connections only to the visible units in the RBM.
\begin{figure}
	\centering
	\includegraphics[clip, trim=0cm 0cm 0cm 0cm,
	width=0.7\linewidth]{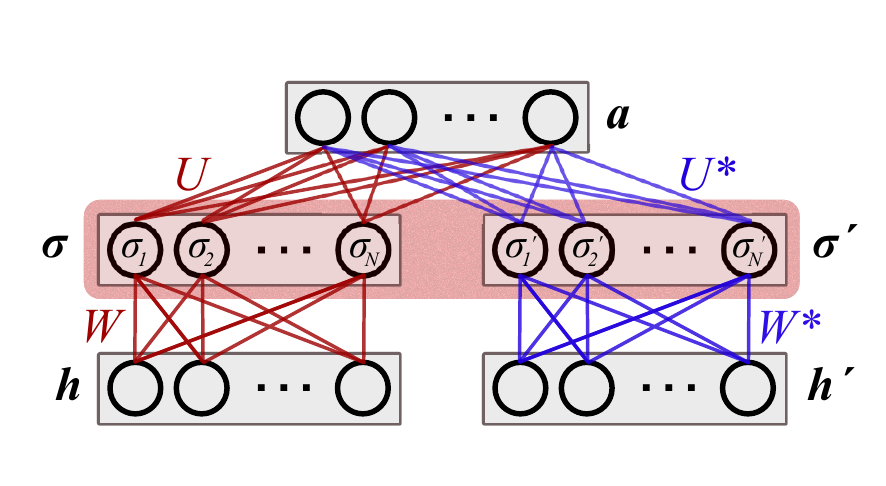}
	\caption{Visualization of the RBM network architecture as proposed by Ref. \cite{torlai_latent_2018} with connections with the weight matrices $U$ and $W$ between the $N$ visible (physical) nodes $\si$ and the ancillary nodes $\ve{a}$  representing the bath, and hidden nodes $\ve{h}$ encoding correlations between individual physical spins. 
		The hidden layers are doubled for the amplitude and phase parts of the network, see Eq.~\eqref{eq:psi_split_real_imag} and all nodes have possible additional bias terms. 
		The nodes ($\si, \si'$)  represent the input of the final density operator.}
	\label{fig:NDO_RBM_architecture} 
\end{figure}
In Fig.~\ref{fig:NDO_RBM_architecture} we visualize the connections between the different layers of nodes. 
The output $P_\ve{X}$ of such a RBM ansatz with parameters $\ve{X} = (\ve{c}^X, \ve{b}^X, \ve{W}^X, \ve{d}^X, \ve{U}^X)$, where only the hidden units $\ve{h}$ are traced out, is then given by
\begin{equation}
	\begin{split}
	P_\ve{X}(\si, \ve{a}) = \sum_{\ve{h}} 
	&e^{\sum_j c_j^X h_j + \sum_i b_i^X \sigma_i + \sum_{ij} W_{ij}^X \sigma_i h_j} \\
	&e^{\sum_k d_k^X a_k + \sum_{ik} U_{ik}^X \sigma_i a_k }  \; .
	\end{split}
\end{equation}
The sums over $\ve{h}$ and $\ve{a}$ in Eq.~\eqref{eq:rho_ansatz} can be carried out 
analytically 
which yields a simple variational form, see Appendix A or Ref.~\cite{torlai_latent_2018}.
In analogy to the hidden layer density $\alpha$ the ancillary layer density $\beta = K / N$ is introduced to control the number of parameters.

\subsection{Dissipative Driven Heisenberg Spin Chain}
\label{sec:dissipative_driven_chain}

Here, as in previous papers Refs.~\cite{kaestle_sampling_2021, prosen_exact_2011} that will be discussed below, 
we consider an anisotropic 
spin-1/2 Heisenberg (XXZ) chain of $N$ spins with open boundary conditions described by the 
Hamilton matrix in the $S^{z}$-basis 
\begin{equation}
	H = J \; \sum_{k=1}^{N-1} \left( \sigma_k^x \sigma_{k+1}^x + \sigma_k^y \sigma_{k+1}^y  + \Delta \; \sigma_k^z \sigma_{k+1}^z \right)\\ \; 
	\label{eq:Heisenberg}
\end{equation}
where $\sigma^\alpha_k$ is a Pauli spin matrix for site $k$, $J$ is a coupling strength 
and $\Delta$ is the anisotropy parameter. 
As far as the Lindblad super-operator is concerned we 
use 
the following $2N$ jump operators, using $\sigma^\pm = \frac{1}{2} (\sigma^x \pm i \sigma^y) $,
\begin{align}
\{ L^{\dagger}_{k} \} &=
\{
\sqrt{\gamma^{+}_{1}} \sigma^{+}_{1},
\ldots,
\sqrt{\gamma^{+}_{N}} \sigma^{+}_{N},\;
\sqrt{\gamma^{-}_{1}} \sigma^{-}_{1},
\ldots,
\sqrt{\gamma^{-}_{N}} \sigma^{-}_{N}
\} \; .
\end{align}

NDO based on the RBM were initially 
tested on systems with homogenous 
dissipation 
\cite{hartmann_neural-network_2019, 
	nagy_variational_2019, 
	vicentini_variational_2019, 
	yoshioka_constructing_2019},
i.e. $\gamma^{\alpha}_{i} = \gamma\;, \forall i$. 
For a spin chain that is asymmetrically driven at the boundaries 
e.g. $\gamma_{1}^{\alpha} \ne \gamma_{N}^{\alpha}$, 
the authors of Ref.~\cite{kaestle_sampling_2021} reported difficulties in obtaining accurate results 
using the above mentioned approach for the Monte-Carlo optimization of the parameters of a similar RBM density operator. 
Therefore they proposed a modified Metropolis sampling procedure, treating the boundary spins exactly while sampling only the bulk of the chain, which leads to more accurate results. 
However, the authors test their approach only for local quantities,  i.e. the site-dependent magnetization. 
In the present work we show that local quantities, being related to the real diagonal elements of the density operator only, are less sensitive to inaccuracies of $\rho$ itself. 
In other words, accurate values for these quantities are not necessarily associated with a high fidelity of $\rho$.

To the best of our knowledge 
Ref.~\cite{kaestle_sampling_2021} is the only work so far addressing \emph{asymmetric} open quantum system treated within NDO, investigating an isotropic Heisenberg chain 
with $\gamma_{i}^{\alpha}=\gamma=0.2$ for all bulk sites $i=2,\ldots,N-1$ and
$\gamma_{1}^{\alpha} = \gamma_{N}^{-\alpha}$, $\gamma^{+}_{1} = (1+\delta) \gamma^{-}_{1}$, $\gamma_{1}^{-} = 0.2$.
We refer to this as model A, to be distinguished with model B introduced below. 
With a \emph{bias} coefficient $\delta = 0.05$, an asymmetric driving is obtained which produces a (non conserved) current.  
In order to compare with the results of Ref.~\cite{kaestle_sampling_2021} we now choose the same 
hidden layer density $\alpha = \beta = 1$, set 
$J$ equal to $0.105$ (which in this case is equal to $2 \gamma_1^+ / 4$) and 
$\Delta= 1$ 
and a sample size of 10000 to reduce the variance.

\begin{figure}[h!]
	\centering
	\includegraphics[clip, trim=0cm 0cm 0cm 0cm,
	width=1\linewidth]{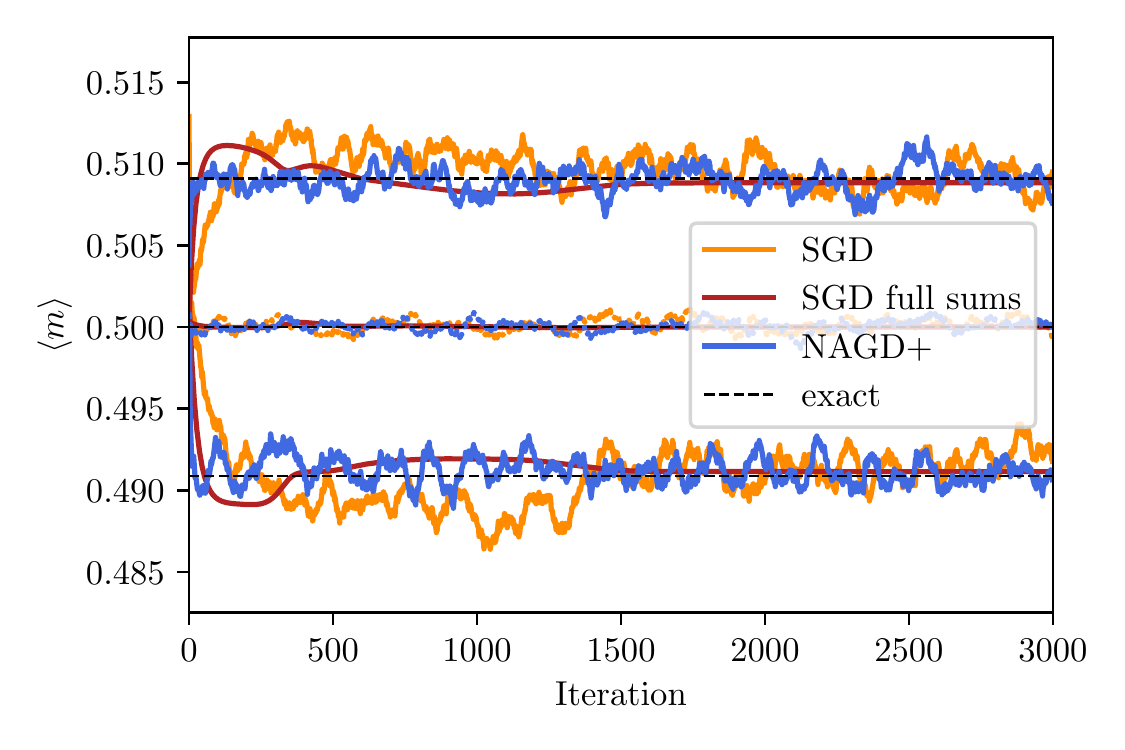}
	\includegraphics[clip, trim=0.332cm 0cm 0.02cm 0cm,
	width=1\linewidth]{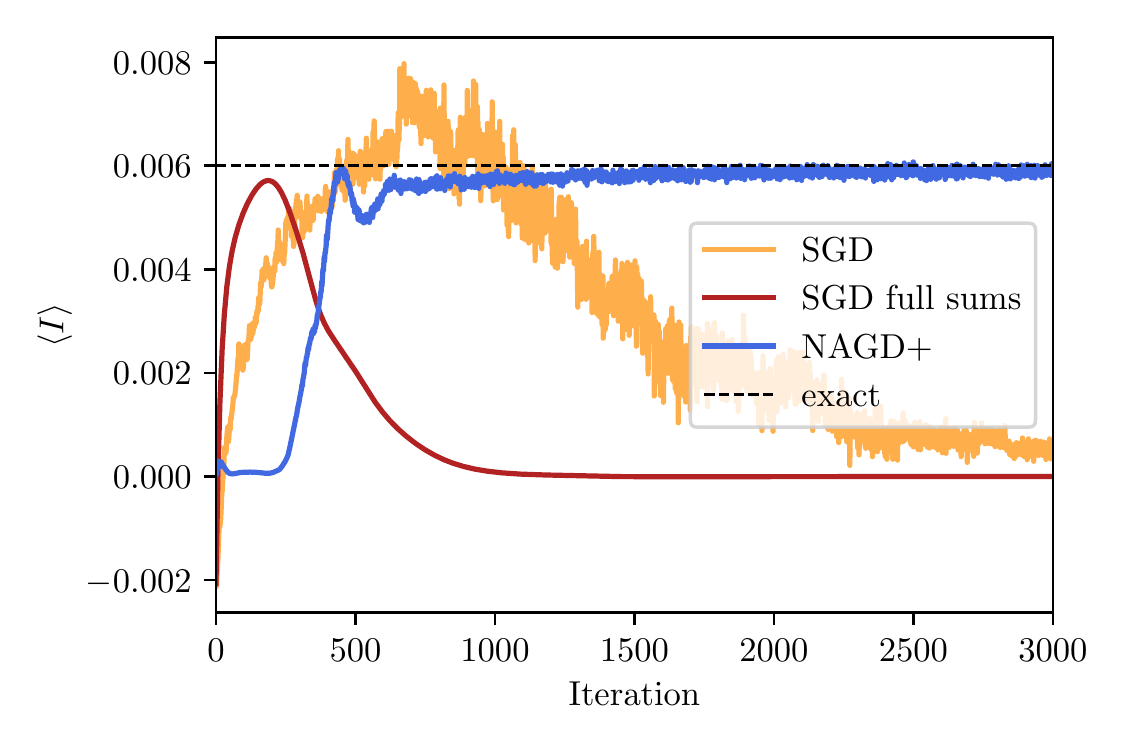}
	\includegraphics[clip, trim=-0.025cm 0cm -0.3cm 0cm,
	width=1\linewidth]{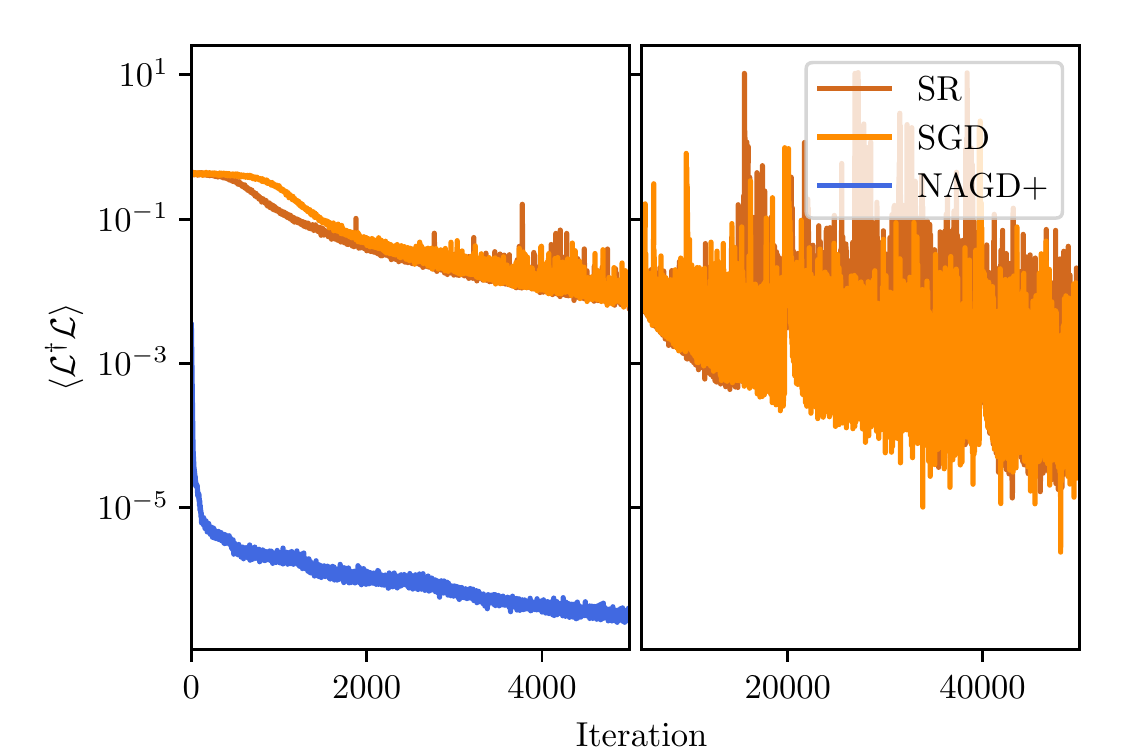}
	\caption{Convergence behaviour of a boundary driven dissipative Heisenberg chain (model A) for $6$ spins using the conventional SGD algorithm with Monte-Carlo samples, SGD with full sums and the presented NAGD+ algorithm compared to exact diagonalization result.  
		Top: Magnetization of the first and last spin (top and bottom lines) as well as the mean bulk magnetization (middle line), running average over 50 iterations. 
		Middle: Spin current $I_{12}$ from a boundary site to its neighbour in the bulk.  
		Bottom: Convergence of the cost function using SGD, SR and the NAGD+ algorithms.
	}
	\label{fig:magnetization_sgd_vs_nagd}
	\hfill
\end{figure}
In Fig.~\ref{fig:magnetization_sgd_vs_nagd} 
we plot the convergence of the boundary  and bulk magnetization for a chain of $6$ spins as a function of the number of iterations. 
Results are obtained by plain SGD update steps, as was done in Ref.~\cite{kaestle_sampling_2021}, both by carrying out the full (external) sums in 
Eqs. (\ref{eq:cost}) and (\ref{eq:observable}) as well as by sampling with a local spin flip Monte-Carlo update. 
The exact diagonalization result is shown for comparison. 
In addition, we show results obtained by our improved Nesterov accelerated gradient descent (NAGD+) approach discussed below in Sec.~\ref{sec:nagd} and \ref{sec:mcmcSampler} and also the progression of the cost function using the SR method which generates a similar result.  
As one can see, the magnetisation converges quite well in all cases 
and the final expectation values have an error of $0.26 \%$ compared to exact results. 
In Fig.~\ref{fig:magnetization_sgd_vs_nagd} (middle panel) we display 
the spin current from first to second site, 
defined as
\footnote{The actual definition includes a factor $J$, which we omit for simplicity.}
\begin{equation}
	\hat{I}_{jk} = i \left( \sigma^+_j \sigma^-_{k} - \sigma^-_j \sigma^+_{k}  \right) \; 
	\label{eq:spinCurrent}
\end{equation}
between spins $j$ and $k$. 
This figure clearly shows that, although the magnetizations 
converge quickly 
to the exact values, 
the spin current converges to a wrong value using SGD both with  
exact summation and MC sampling. 
As we will discuss below, this is due to the fact that the current is related to the imaginary part of the off-diagonal elements of $\rho$, which, being much smaller than the diagonal terms, are obtained much less accurately. 
These shortcomings are not improved by using the more sophisticated SR approach, 
as can be seen in the bottom panel of Fig.~\ref{fig:magnetization_sgd_vs_nagd} where both SR and SGD seem to have difficulty optimizing the model, while the NAGD+ algorithm finds an orders of magnitude lower cost function in a fraction of the iteration steps used for the other two algorithms.  
The learning rates were chosen as $0.01$ for SR and $0.05$ for SGD, and a diagonal shift of $0.1$ in the $S$ matrix was necessary in the SR optimizer for a stable optimization. In all MC runs a sample size of $2000$ was used and the parameters were initialized following a normal distribution around $0$ with $\sigma^2 = 0.01$.

\begin{figure}[h!]
	\label{fig:rho0_vs_rhoRBM_A}
	\centering
	\subfloat[Exact result for 6 spins 		\label{fig:rho0_A}]{
		\centering
		\includegraphics[clip, trim=0cm 1.2cm 0cm 1.2cm, 
		width=\linewidth]
		{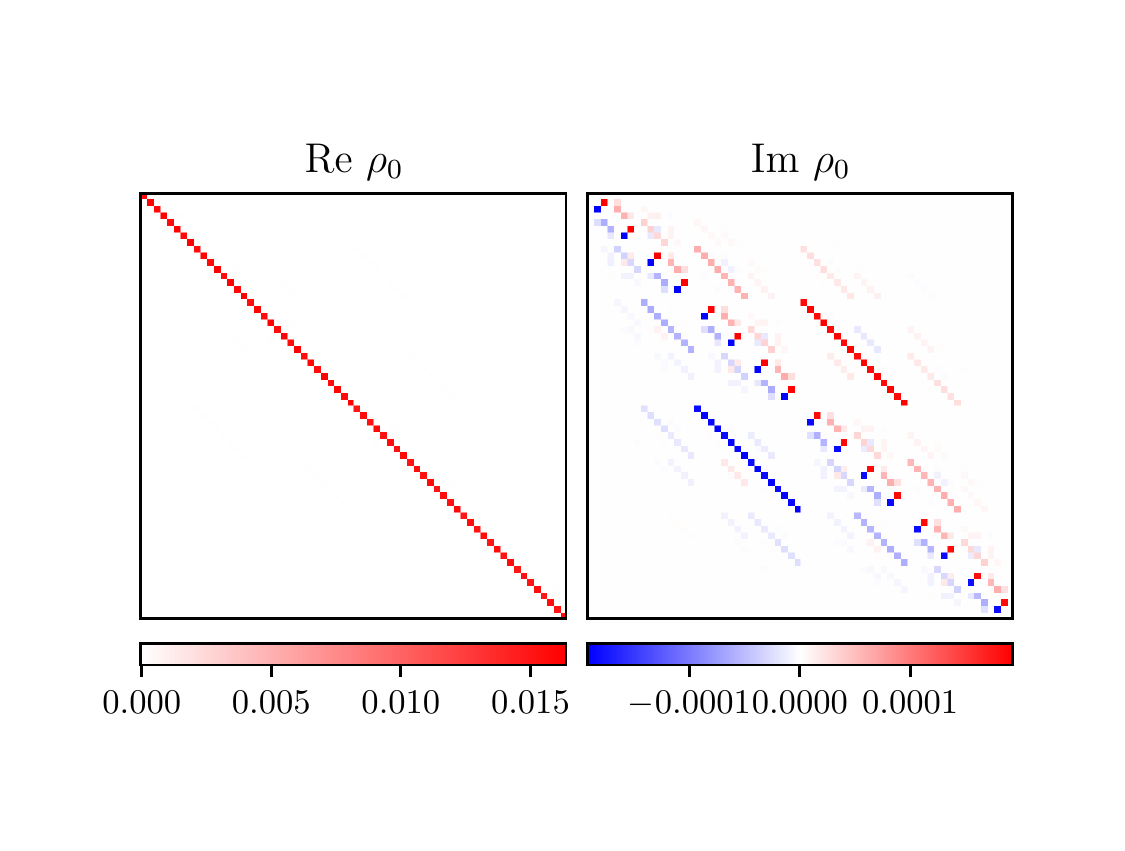}
	}
	\hfill
	\subfloat[Approximation using conventional SR 		\label{fig:rho_rbm_A}]{
		\centering
		\includegraphics[clip, trim=0cm 1.2cm 0cm 1.2cm,
		width=\linewidth]
		{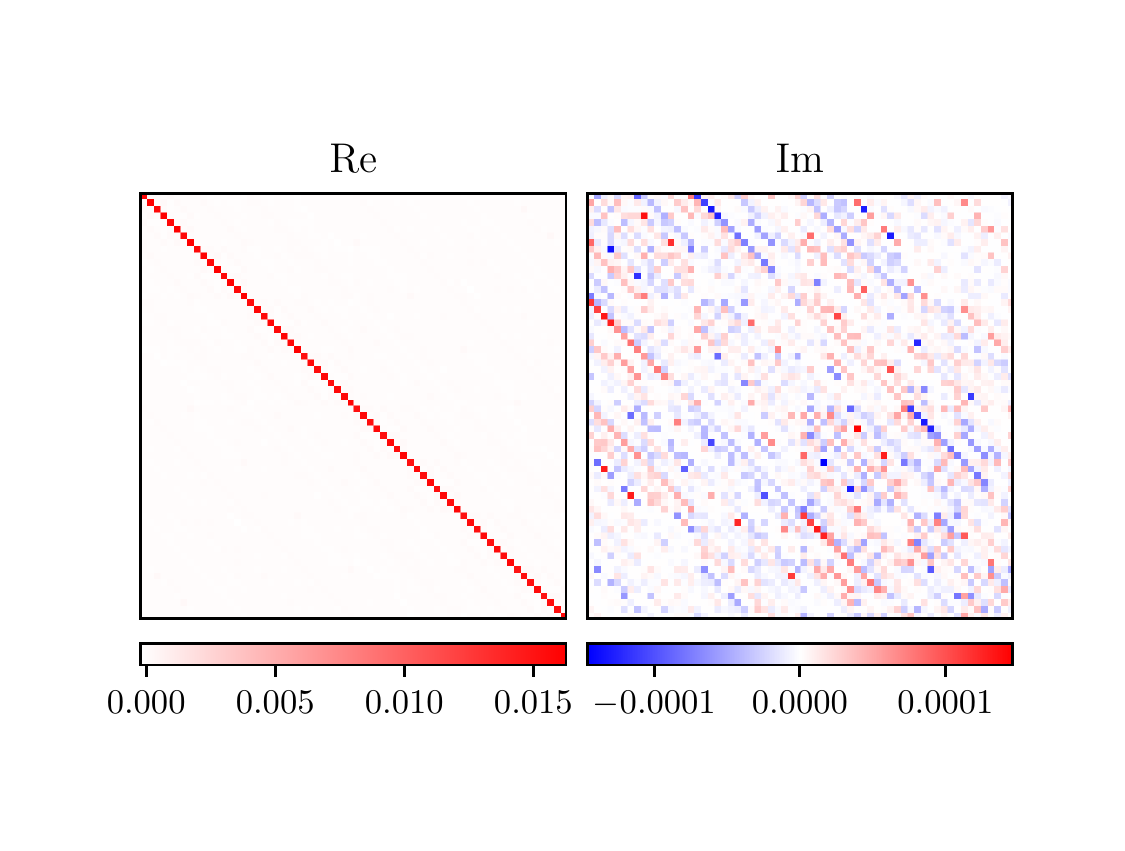}
	}
	\hfill
	\subfloat[Approximation using our optimization method NAGD+ \label{fig:rho_rbm_nagd_A}]{
		\centering
		\includegraphics[clip, trim=0cm 1.2cm 0cm 1.2cm,
		width=\linewidth]
		{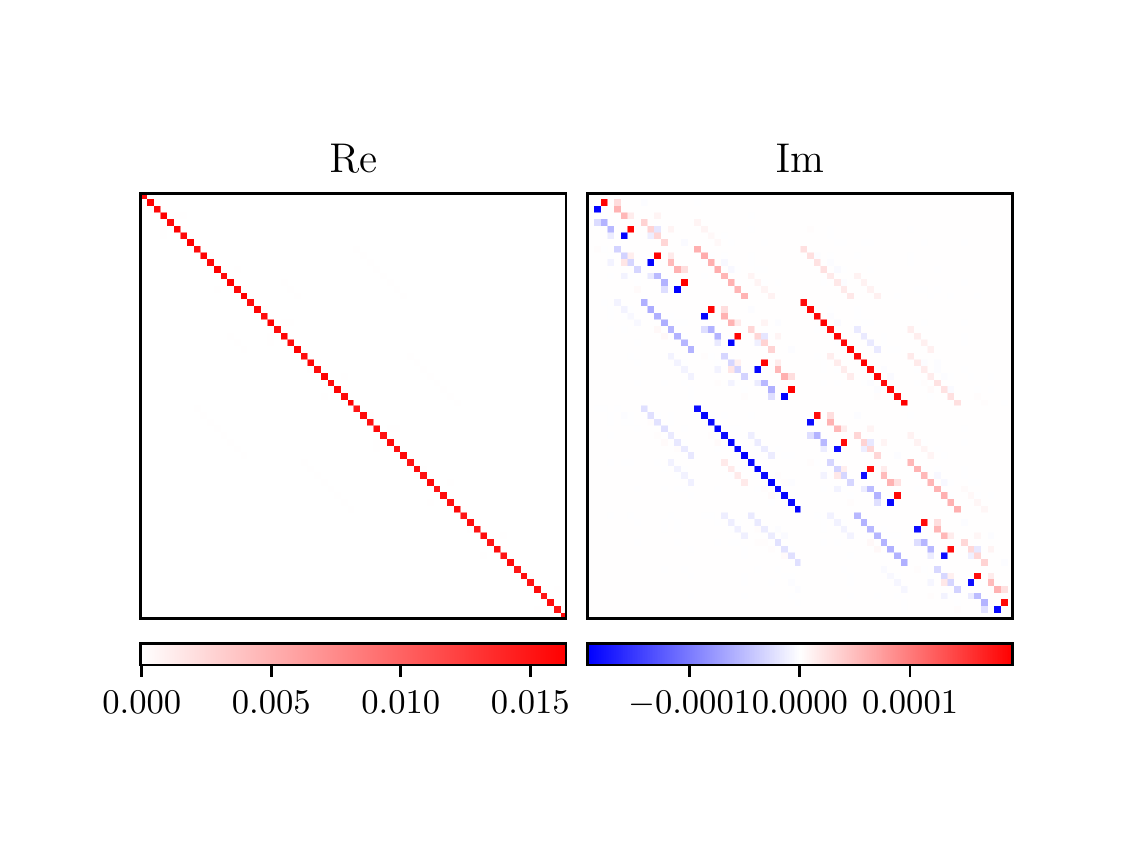}
	}
	\caption{	\label{fig:rho0_vs_rhoRBMA}Real and imaginary parts of the non-equilibrium steady state density matrix of a boundary driven dissipative Heisenberg chain (model A) with 6 sites. (a) exact result, (b) NDO optimized with conventional SR after 50000 iterations and (c) using the proposed NAGD+ algorithm after 3000 iterations. }
\end{figure}

A measure for the accuracy of $\rho$ with respect to the exact density matrix  $\rho_{0}$ 
obtained by exact diagonalization 
is provided by the fidelity $F(\rho, \rho_{0})$ \cite{lidarLectureNotesOnOpenQuantumSystems} defined as 
\begin{equation}
	\label{eq:fidelity}
	F(\rho, \rho_{0}) = \text{Tr} \sqrt{\sqrt{\rho} \; \rho_{0} \sqrt{\rho}}\; ,
\end{equation}
whereby $F=1$ corresponds to an exact match. 
In Fig.~\ref{fig:rho0_vs_rhoRBMA} the density matrix is depicted for model A. 
The density matrix
obtained by SGD after $3000$ iterations yields a value of only $F=0.7$. 
The density matrix obtained by SR with significantly more iterations (50000) is depicted in Fig.~\ref{fig:rho_rbm_A} and reproduces the real part  almost correctly, but the imaginary part is still completely wrong, as one can tell from the comparison with the exact density matrix shown in Fig.~\ref{fig:rho0_A}. 
Nevertheless, the fidelity is already $F=0.9995$ 
because it is dominated by the much larger  real part.
The current, however, is determined by the imaginary part 
of the density matrix and the error of the current remains above $90~\%$. 
This tells us 
that the fidelity is not always a suitable measure for the quality of the approximation. 
The behaviour is completely different when using NAGD+. 
In this case we achieve a fidelity $F>0.999999$ already after 3000 iterations. 
The corresponding density matrix is depicted  in Fig.~\ref{fig:rho_rbm_nagd_A}. 
We see that real and imaginary parts are obtained reliably and consequently we also find 
a good agreement for the current with respect to the exact result (Fig.~\ref{fig:magnetization_sgd_vs_nagd} middle panel).

\begin{figure}[h!]
	\centering
	\subfloat[Exact $\rho_0$ vs SR result $\rho$ for 4 spins 		\label{fig:rho0_l4}]{	
		\centering
		\includegraphics[clip, trim=0cm 3cm 0cm 2.5cm, 
		width=\linewidth]
		{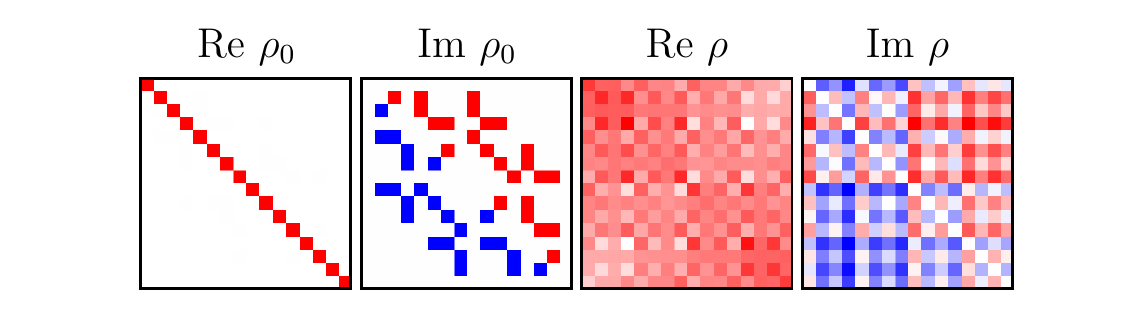}
	}
	\hfill
	\subfloat[Exact result for 6 spins 		\label{fig:rho0}]{	
		\centering
		\includegraphics[clip, trim=0cm 1.2cm 0cm 1.2cm, 
		width=\linewidth]
		{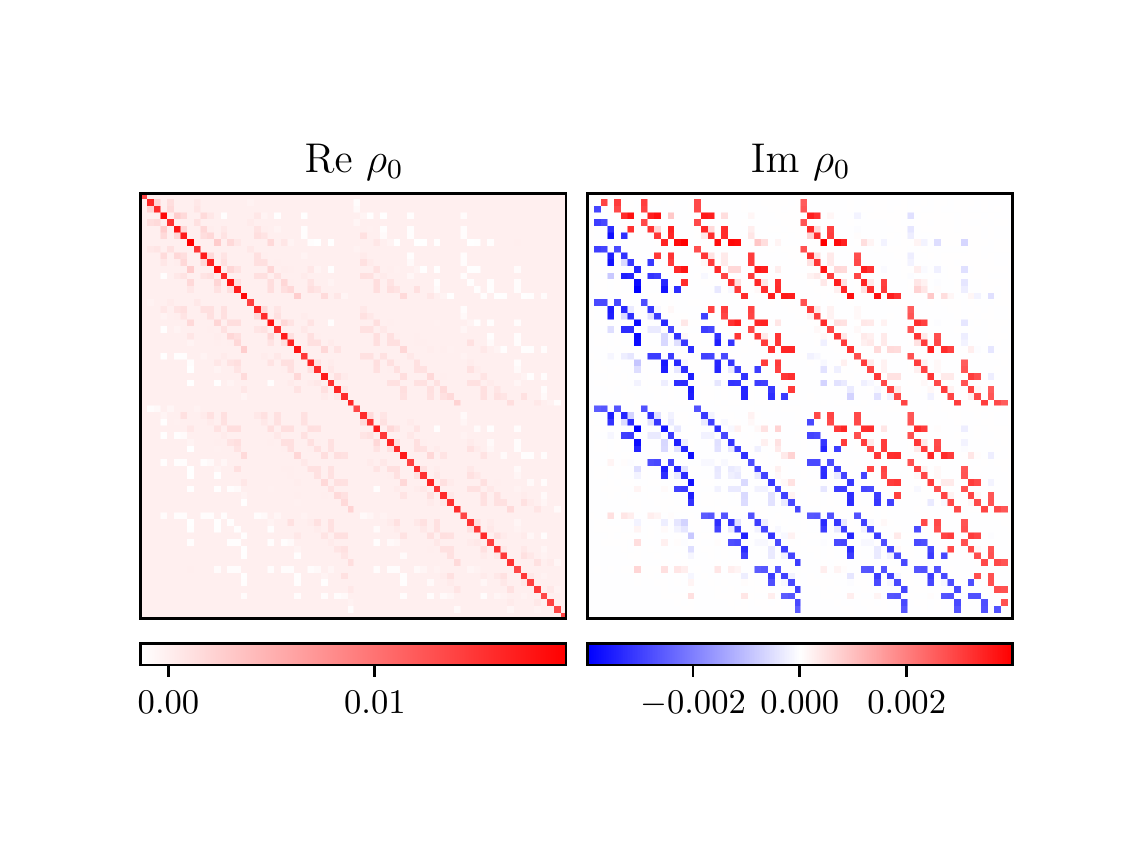}
	}
	\hfill
	\subfloat[Approximation using conventional SR 		\label{fig:rho_rbm}]{
		\centering
		\includegraphics[clip, trim=0cm 1.2cm 0cm 1.2cm,
		width=\linewidth]
		{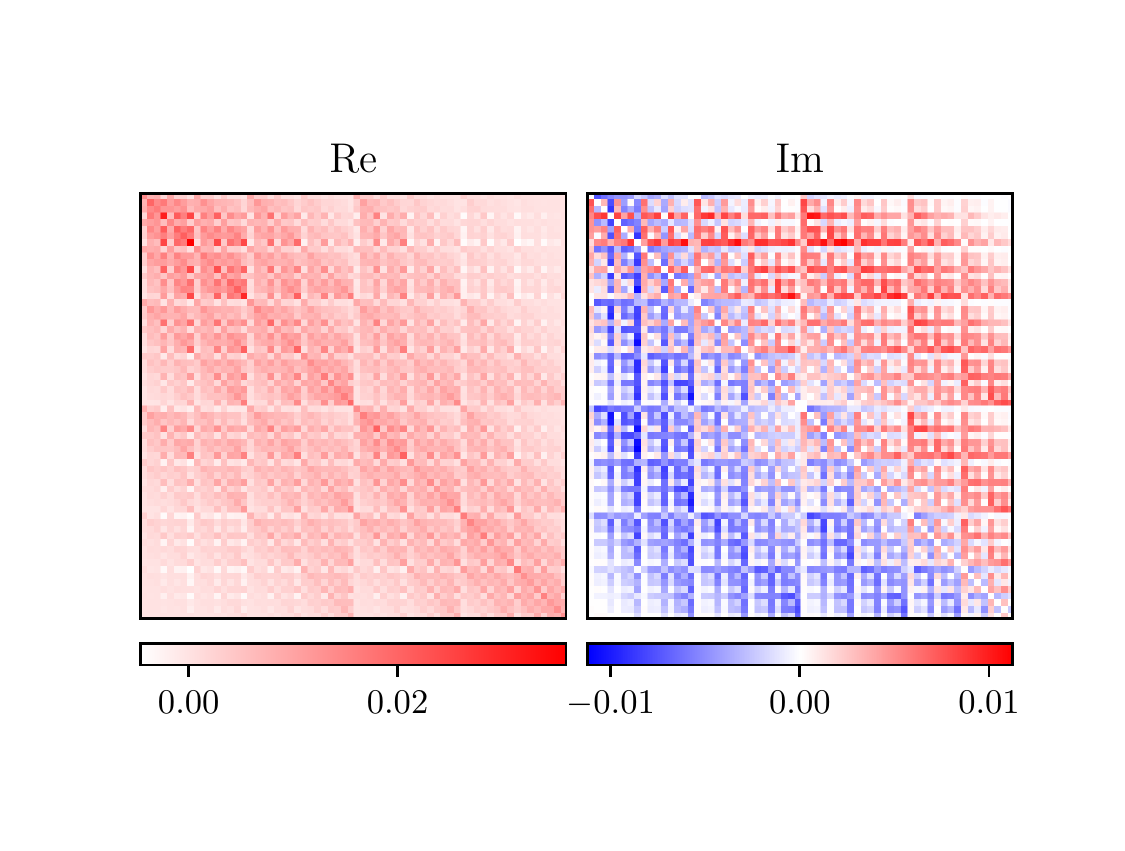}
	}
	\hfill
	\subfloat[Approximation using our optimization method NAGD+ 		\label{fig:rho_rbm_nagd}]{
		\centering
		\includegraphics[clip, trim=0cm 1.2cm 0cm 1.2cm,
		width=\linewidth]
		{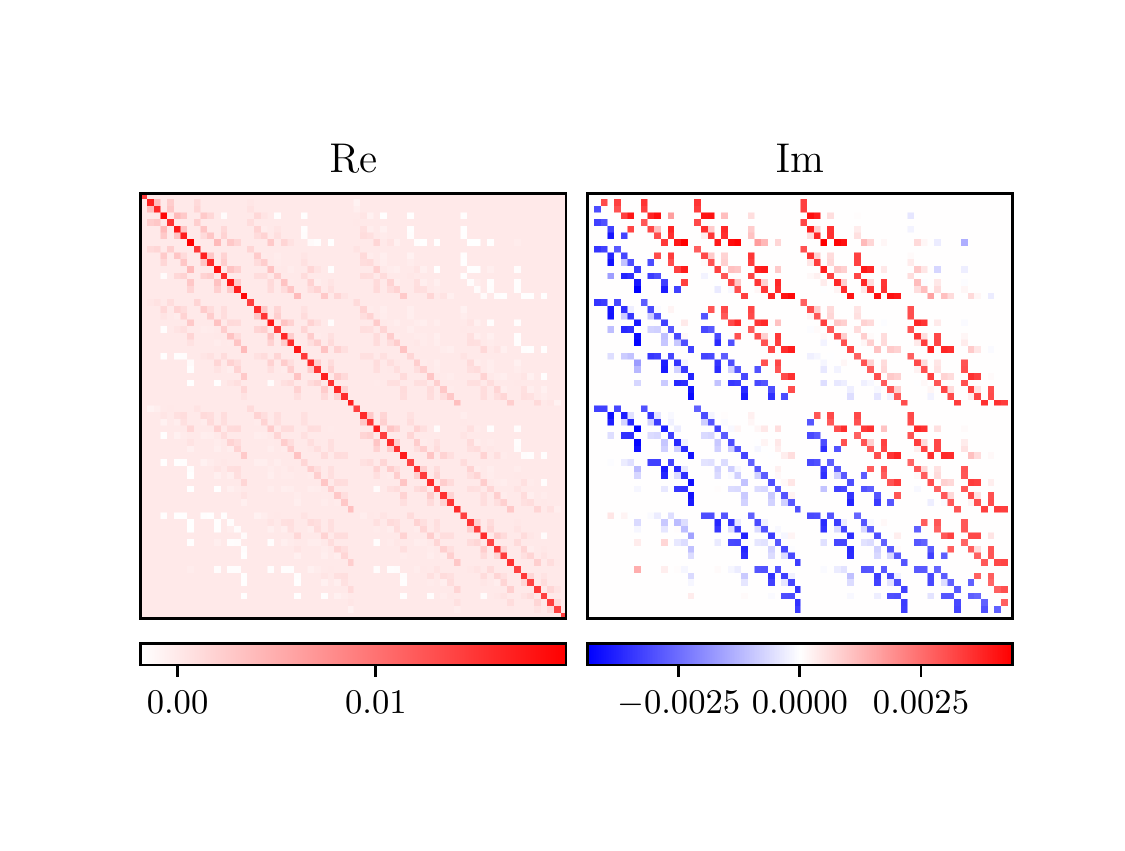}
	}
	\caption{Real and imaginary parts of the non-equilibrium steady state density matrix of a boundary driven Heisenberg chain (model B). 
	For a 4 site chain (a) the results obtained by exact diagonalization are  compared  with the result of the NDO ansatz optimized with conventional SR. 
	(b-d) show the density matrix for a 6-site chain: (b) exact result, (c) SR and (d) NAGD+ results.}
	\label{fig:rho0_vs_rhoRBM}
\end{figure}

We now consider another current-carrying open spin system. This consists of an isotropic ($\Delta = 1$) Heisenberg chain with just one jump operator 
$L^+ = \sqrt{\gamma} \sigma_1^+$ 
on the first and one 
$L^- = \sqrt{\gamma} \sigma_N^-$ 
on the last site, 
corresponding to $\gamma_{1}^{+} = \gamma = \gamma_{N}^{-}= \gamma$ and all other $\gamma_{i}^{\alpha}=0$. 
This system (model B) can even be treated analytically \cite{prosen_exact_2011}. 
The asymmetry resembles a bias voltage across the chain which 
again induces
a current. 
We take $\Delta = J = 1$ and set a dissipation strength of $\gamma = 0.2$. 
For the hidden node densities we pick $\alpha = \beta = 2$. 
%
This model has the same basic structure of the NESS density matrix as in the previous results (Fig.~\ref{fig:rho0_A}) 
with a predominantly diagonal real part and sparse imaginary values that are orders of magnitude smaller, as seen in Fig.~\ref{fig:rho0}. 
In Fig.~\ref{fig:rho0_vs_rhoRBM} the final density matrix approximation obtained after 10000 iterations 
is compared to the exact one. 
Using the established stochastic reconfiguration methods 
the steady-state for these systems cannot be found even when the sums are computed exactly, reaching a fidelity of only $F=0.6$ 
after converging at a very slow rate, and it barely improves after another 40000 iterations to reach $F=0.8$. 
Looking at another example of just 4 spins with a smaller dissipation $\gamma = 0.04$ the SR method gets completely stuck achieving a fidelity below $0.5$ as shown in Fig.~\ref{fig:rho0_l4} on the right. We found this behaviour for all random initial parameters we tried.


\subsection{Advanced Optimization Algorithm to Find a High Fidelity Steady State}
\label{sec:nagd}
Above we have already compared results obtained with state of the art 
ML 
approaches for non-equilibrium systems with NAGD+, an advanced optimization method, which we describe in detail below. 
We combine backtracking Nesterov accelerated gradient descent (NAGD)
with dynamic scaling of individual directions (preconditioning) known from AdaBelief \cite{zhuang_adabelief_2020}
to obtain an optimization scheme that converges significantly faster and 
reaches orders of magnitude lower residues than the conventional stochastic reconfiguration approach. 

Starting from a parameter set $x_t$ at iteration $t$, Nesterov gradient descent  
introduces an intermediate step $y_t$ to extrapolate in the direction of the previous update 
scaled by a factor $\gamma_t$. 
The gradient is then computed at that point and the parameters $x$ are updated with a step-size $\eta_t$
\begin{align}
\label{ex:aux0}
y_t &= x_t + \gamma_t \; (x_t - x_{t-1}) \\
x_{t+1} &= y_t - \eta_t \; \nabla f(y_t)  
\label{ex:aux1}
\; .
\end{align}
The extrapolation parameter $\gamma_t$ is usually set to a fixed value (in our case $0.9$) or dynamically increased $\gamma_t = t/(t+3)$, which in our experience converges faster but can lead to instabilities.  
The step size $\eta_{t}$ is often considered arbitrary,  but can be motivated by considering the descent lemma, 
which provides an upper bound for functions $f$ with Lipschitz continuous gradients (see \cite{BeckFirstOrderMethodsInOptimization} Sec.~5.1.1.)
\begin{align}
\label{eq:sgdLemma}
f(x)
\leq \bar{f}(x) :=
f(y_t) + \langle \nabla f(y_t),  x - y_t \rangle + \frac{\bar{L}}{2} \left\|x - y_t \right\|^2
\; .
\end{align}
The idea behind 
gradient descent from an optimization perspective is the following. 
The cost function $f(x)$ is approximated by the upper bound  $\bar{f}(x)$, for which  the minimum w.r.t. $x$, i.e. 
 \begin{align}
	 x_{t+1} 
	 = \underset{x}{\text{argmin}} 
	 \bar{f}(x)
	 \label{eq:sgdLemma-update}
	 \; 
 \end{align}
is determined. 
This yields the update step in Eq.~\eqref{ex:aux1} with the step size $\eta_{t} = 1/\bar{L}$
being the inverse local Lipschitz constant $\bar{L}$. 
The latter is initially approximated by a suitable value $L_{0}$. If the inequality in Eq.~\eqref{eq:sgdLemma} is violated, it means that the assumed Lipschitz constant $\bar{L}$ is too small or the step size $\eta_{t}$ is too large.
In a so-called backtracking step $L_{0}$ is repeatedly increased (f.e. multiplied by a factor of 2) until the inequality in Eq.~\eqref{eq:sgdLemma} is satisfied.  
Finally, the parameters can then be updated with this adjusted step-size and the next iteration 
starts again with Eq.~\eqref{ex:aux0}. 
Since the Lipschitz constant is a local quantity, it will change during the optimization and we therefore also try to decrease $L_{0}$ (f.e. divide by two) whenever the inequality is immediately satisfied. 
The Monte-Carlo sampling makes up the largest part of the computation time, so the backtracking part of the algorithm doesn't require significantly more computing resources, as it just means reevaluating the network with different parameters without drawing new samples. 

Updates of parameters can be scaled individually by dividing by the exponential moving average of the variance of subsequent gradients, as it is done in the AdaBelief optimizer \cite{zhuang_adabelief_2020}. 
This leads to larger steps for parameters where the variance is small, hence the \emph{belief} in the prediction of the gradient is strong, and smaller steps otherwise. 
In terms of the descent lemma this preconditioning just means introducing a different metric in Eq.~\eqref{eq:sgdLemma}. 

Some additional modifications can further speed up convergence: 
Especially when using exact sums, the model can get stuck in local minima early on, which can often be avoided with some small additive noise on the parameters. 
%
Also, a small number of initialization steps minimizing the mean squared error $|| \rho - \mathbb{1} ||^2 / 2^{2N} $ of our ansatz function 
with respect to the unit matrix 
can also speed up the convergence significantly. 
For larger systems the evaluation of all matrix elements would be too demanding, 
so in this case randomly drawn subsamples of $\{\rho_{ij}\}$ can be used to approximate the gradients of the MSE cost function. 
In the results presented for model A we made 10 such steps after initializing the RBM's parameters to normally distributed values with standard deviation $0.01$. 

With this optimization 
algorithm we are able to optimize the RBM ansatz, finding a high-fidelity steady state representation which also produces correct results for the current. 
Applied to the 6 spin Heisenberg chain with just two jump operators 
(model B) a density matrix with almost indiscernible deviations from the exact result is reached as depicted in Fig.~\ref{fig:rho_rbm_nagd} with a fidelity of over $0.99$. 

\subsection{Monte-Carlo Sampling}
\label{sec:mcmcSampler}
%
The algorithm described in the previous section works well with exact sums in equation \eqref{eq:cost}, but this becomes too computationally intensive for larger systems. So it is common practice to replace the outer sum by  Monte-Carlo sampling.
We will discuss in this section that standard MC sampling 
has a couple of problems and how they can be overcome. 
In Ref.~\cite{kaestle_sampling_2021} it was reported that conventional Monte-Carlo sampling is not suitable  for asymmetrically driven systems. 
The reason for this can be understood by looking at the exact stationary density matrix in Fig.~\ref{fig:rho0}, whose absolute values squared give the probability of a sample. 
Close to convergence to the steady-state 
the density matrix will be sparse, with predominantly diagonal real parts and only some imaginary elements being non-zero.  
Another difficulty is that real and imaginary parts differ by several orders of magnitude 
This has the effect that, when starting from a diagonal sample, a random new proposal configuration is very unlikely to be accepted. 

A way to 
overcome this issue is to \emph{reweight} the probabilities. 
The cost function defined in Eq.~(\ref{eq:cost}) can be rewritten using a simplified notation for the configurations $x = (\si, \si')$ and the \emph{local} quantity 
\begin{align}
{\cal C }(x) =\left| \sum_{x'} \mathcal{L}_{xx'} \frac{\rho({x'})}{\rho(x) } \right|^2
\end{align}
as
\begin{equation}
\begin{split}
	C 
	&=\frac{\sum_x |\rho(x)|^2 \; {\cal C}(x)}{\sum_x |\rho(x)|^2}\\
	&=\frac{\sum_x |\rho(x)|^{2\beta} \; |\rho(x)|^{2-2\beta} \; {\cal C }(x)}{\sum_x |\rho(x)|^{2\beta} \; |\rho(x)|^{2-2\beta}} \\
	&= \frac{    \left\langle 		|\rho(x)|^{2 - 2\beta} \;  {\cal C}(x) \right \rangle_{\tilde{p}(x)} }
	{\left\langle |\rho(x)|^{2 - 2\beta} \right \rangle_{\tilde{p}(x)} }\; .
\end{split}
\end{equation}
%
These expectation values are now computed based on a new probability distribution $\tilde{p}(x)  = |\rho(x)|^{2\beta} / \tilde{Z}$ where the additional normalization $\tilde{Z}$ 
in  numerator and denominator cancel. 
Now we estimate the cost function by drawing Monte-Carlo samples from the new probability distribution. 
We use the same samples for the numerator and the denominator. 
Note that the new probability distribution varies less between different configurations, allowing for simpler random walk in configuration space at a sort of  \emph{higher simulation temperature (for $\beta < 1$)}. 
Compared to the exact treatment of a part of the configuration space, as done in \cite{kaestle_sampling_2021} to capture asymmetric systems more efficiently, our method has the advantage that the asymmetries are not limited to a small number of sites. 
Using the reweighting technique together with our optimization algorithm we are now able find high fidelity steady states even without using full sums. 

\begin{figure}[h!]
	\centering
	\includegraphics[clip, trim=0cm 1cm 0cm 1cm,
	width=0.6\linewidth]{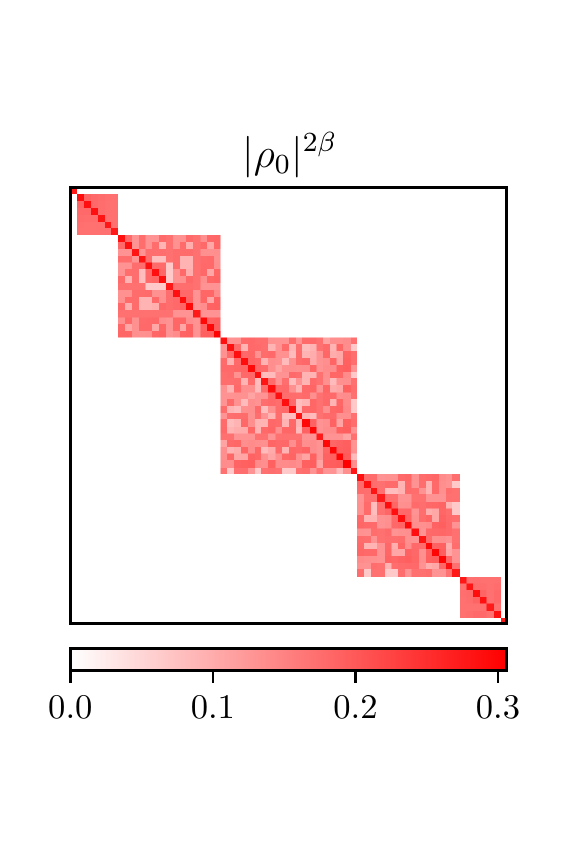}
	\caption{Exact NESS $\rho_0$ of a boundary driven open Heisenberg chain of 6 spins after being reordered according to total $S^z$ and reweighted with $\beta= 0.15$ as the new probability distribution for efficient Monte-Carlo sampling. }
	\label{fig:rho0_2beta}
	\hfill
\end{figure}

There is another simple improvement we can do to speed up convergence.  
{The time evolution  of the density matrix
\begin{align*}
\bra{\sigma}\dot \rho \ket{\sigma'}  &= \bra{\sigma}L \rho \ket{\sigma'} = \sum_{\tau,\tau'}L_{\sigma\sigma',\tau,\tau'} \rho_{\tau,\tau'}
\end{align*}
has a special symmetry that concerns 
the total $S^{z}$-spin  of a configuration $\si$, which we define  by $S^{z}(\si) = \sum_{i} \sigma_{i}$.
In addition, we define the difference of the total $S^{z}$-spin  $\Delta S^{z}(\si,\si') = S^{z}(\si)-S^{z}(\si')$ of the configuration pair $\si,\si'$. Then, due to the particular structure of the Lindblad operator,  $L_{\sigma\sigma',\tau,\tau'}$ is non-zero only for $\Delta S(\si,\si') = \Delta S(\ve{\tau},\ve{\tau}')$.
 Therefore, during the time evolution, the parts of $\rho(\si,\si')$ that belong to different $\Delta S^{z}$  values do not couple. Put differently, if we start with a density matrix that has non-zero elements only for configuration pairs $(\si,\si')$ that have the same 
 $\Delta S^{z}(\si,\si')$ 
 then this sector is not left during the time evolution and also the converged state belongs to this sector. On the other hand, a matrix 
 $\rho$ that belongs to a sector $\Delta S^{z}\ne 0$ has zero diagonal elements and therefore $\text{tr} \rho = 0$ and, hence, it does not qualify as a density matrix.  Therefore, we merely need to consider the sector $\Delta S^{z}(\si,\si') = 0$.
	}
This has the effect that the steady-state density matrix separates into blocks with equal total spin $S^z$. 
The block structure of the exact steady state density of the previous problem (model B with 6 sites) is depicted
 in Fig.~\ref{fig:rho0_2beta}, where the spin configurations $\si$ are reordered according to the total $S^z$ values and we present the reweighed modulus squared  
 of the density matrix
 $|\rho(\si, \si')|^{2\beta}$ with $\beta = 0.15$.

Since the density matrix ansatz appears mainly as the probability distribution during sampling, 
the easiest way of projecting onto only those subspaces where $S^z$ is conserved is to remove all other configurations from the Markov-chain in the first place.  
We therefore propose a sampling mechanism where we avoid any configurations where $\Delta S^z$ is not conserved. 
To find a new proposal given a valid starting configuration $\boldsymbol x = (\si, \si') = (\sigma_1, ..., \sigma_N, \sigma'_1, ...,\sigma'_N)$
we perform two consecutive moves:
\begin{enumerate}
	\item Choose a  pair of site indices $i,j$ at random and swap either the spins $\sigma_{i}$ and $\sigma_{j}$ or 
the spins $\sigma'_{i}$ and $\sigma'_{j}$ .

\item Choose  a  new pair of indices $k,l$ 
at random and flip  the spins 
$\sigma_{k}\to -\sigma_{k}$ and $\sigma'_{l}\to -\sigma'_{l}$ 
if they have the same value.
\end{enumerate}

\noindent
The first step makes moves within a 
diagonal sub-block of a total $S^z$ whereas the second step moves from one such 
sub-block to another. 
Both steps are ergodic. First of all note that the different configurations $\si$ within one sub-block are related by permutations of the site indices. Secondly, the first step of the moves describes a transposition of the site indices and each permutation can be constructed by a suitable sequence of transpositions. In case of the second step, the total $S^{z}$ is changed by $\pm 2$ and hence all $S^{z}$ values can be reached. Because of the constraint $\sigma_{k}=\sigma'_{l}$ the total $S_{z}$ in both configurations $\si$ and $\si'$ changes in the same way.
In addition, in both steps the proposal probability  is symmetric $p^\text{pr}(\boldsymbol x\to\boldsymbol x') =
p^\text{pr}(\boldsymbol x'\to\boldsymbol x)$,  resulting in unbiased sampling. In both steps the probability is $1/N_{p}$  (where $N_{p}$ is the number of possible index pairs) to choose an index pair and that leads uniquely from configuration $\boldsymbol x$
to $\boldsymbol x'$, where the two configurations can be equal. 
The reverse move is generated by the same index pair, and hence with the same probability.
Ergodicity and unbiasedness   can be easily verified by setting all Monte-Carlo weights to 1 and check for a uniform distribution.

\section{Results of Steady-State Spin Current in Asymmetrically Driven Dissipative XXZ chains}

\begin{figure}[h]
	\centering
	\includegraphics[clip, trim=0cm 0cm 0cm 0cm,
	width=1\linewidth]{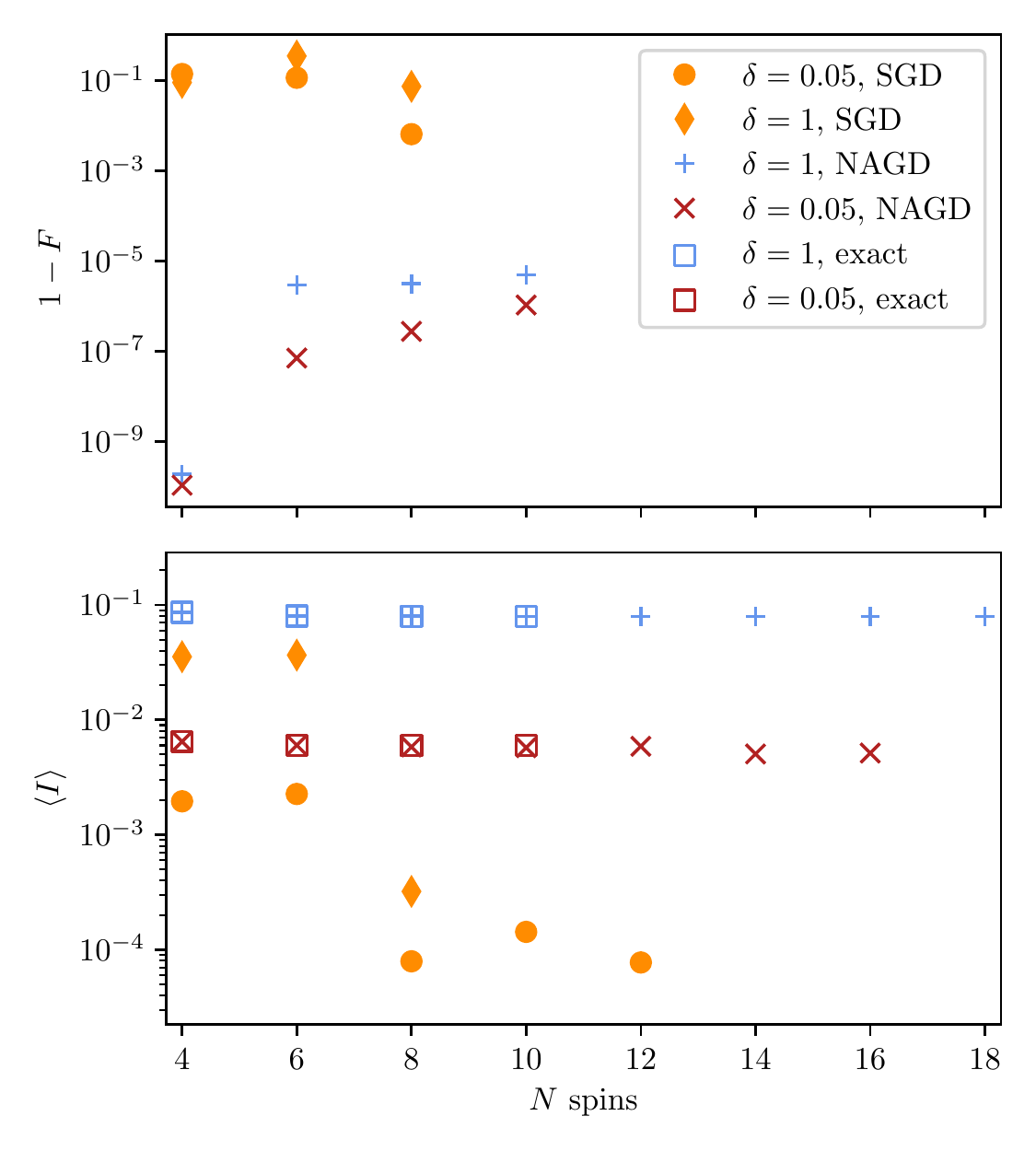}
	\caption{
	Fidelity $F$ (where an exact result was available) and spin current $I$ as a function of the number of spins after 10000 iterations using conventional SGD and NAGD+ optimization for a boundary driven dissipative Heisenberg spin chain (model A) with 5 \% bias and 100 \% bias ($\delta$) as described in the beginning of section~\ref{sec:dissipative_driven_chain}, compared to exact diagonalization.}
	\label{fig:current_A_withLargerBias}
	\hfill
\end{figure}

\begin{figure}[h]
	\centering
	\includegraphics[clip, trim=0cm 0cm 0cm 0cm,
	width=1\linewidth]{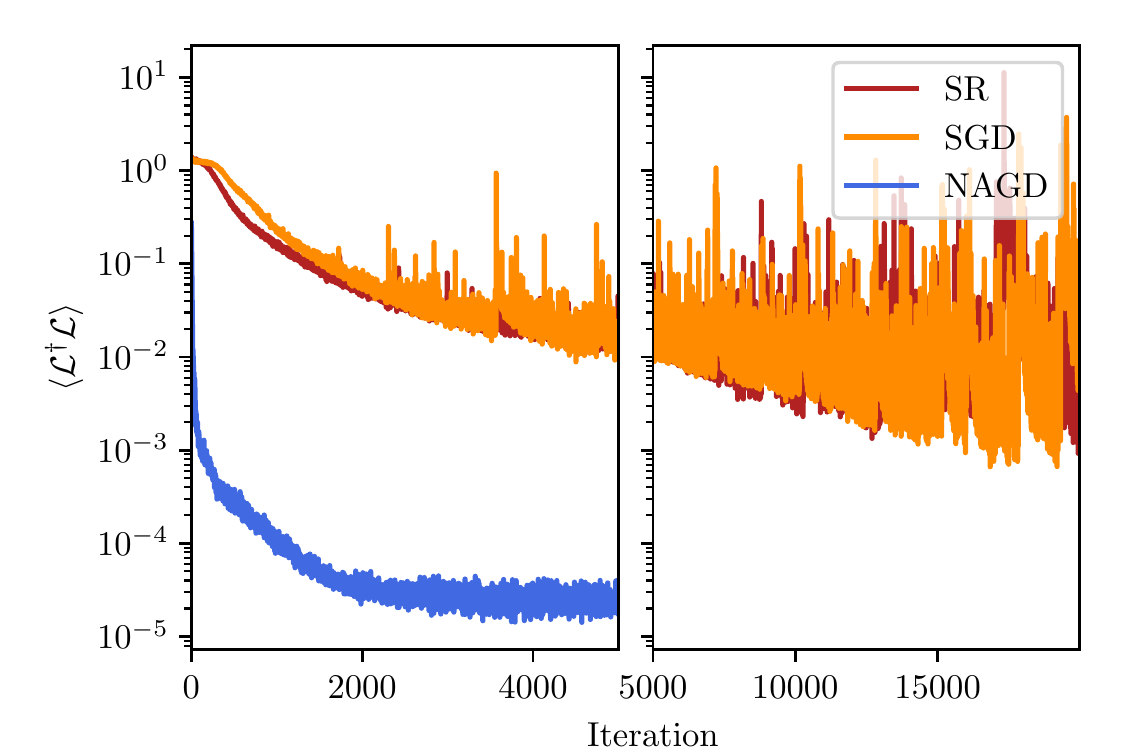}
	\caption{Convergence of $\langle \Li^\dagger\Li \rangle$ using conventional SGD, SR and NAGD+ optimization in a boundary driven dissipative Heisenberg chain (model A) of 10 spins with bias $\delta=1$ as described in the beginning of section~\ref{sec:dissipative_driven_chain}. 
A similar convergence of the different methods is observed as for smaller chains in Fig.~\ref{fig:magnetization_sgd_vs_nagd} (bottom).}

	\label{fig:LdagL_A_LargerBias}
	\hfill
\end{figure}

For the setup from Ref.~\cite{kaestle_sampling_2021} 
(model A) discussed in the beginning of section~\ref{sec:dissipative_driven_chain}, the convergence of the magnetization and spin current in a 6-spin chain is plotted in Fig.~\ref{fig:magnetization_sgd_vs_nagd}. 
The final density matrix using our method has a fidelity of $0.999999$.
In Fig.~\ref{fig:current_A_withLargerBias} the fidelity and current $I_{12}$ for longer chains are plotted for 
{model A} 
with $\delta= 0.05$ ($\gamma^+ = (1+\delta) \gamma^-$ at the boundaries) and also for a larger bias with $\delta= 1$. 
With the NAGD+ optimization we could achieve an excellent agreement with exact diagonalization results with fidelities no worse than 0.99999. Also the spin current matches the exact results. 
All runs were stopped after 10000 iterations, as even though the results using SGD were not completely converged, they were improving at a rate so slow that after double the time the results were not much improved (see Fig.~\ref{fig:LdagL_A_LargerBias}). 
%

\begin{figure}
	\centering
	\includegraphics[clip, trim=0cm 0cm 0cm 0cm,
	width=1\linewidth]{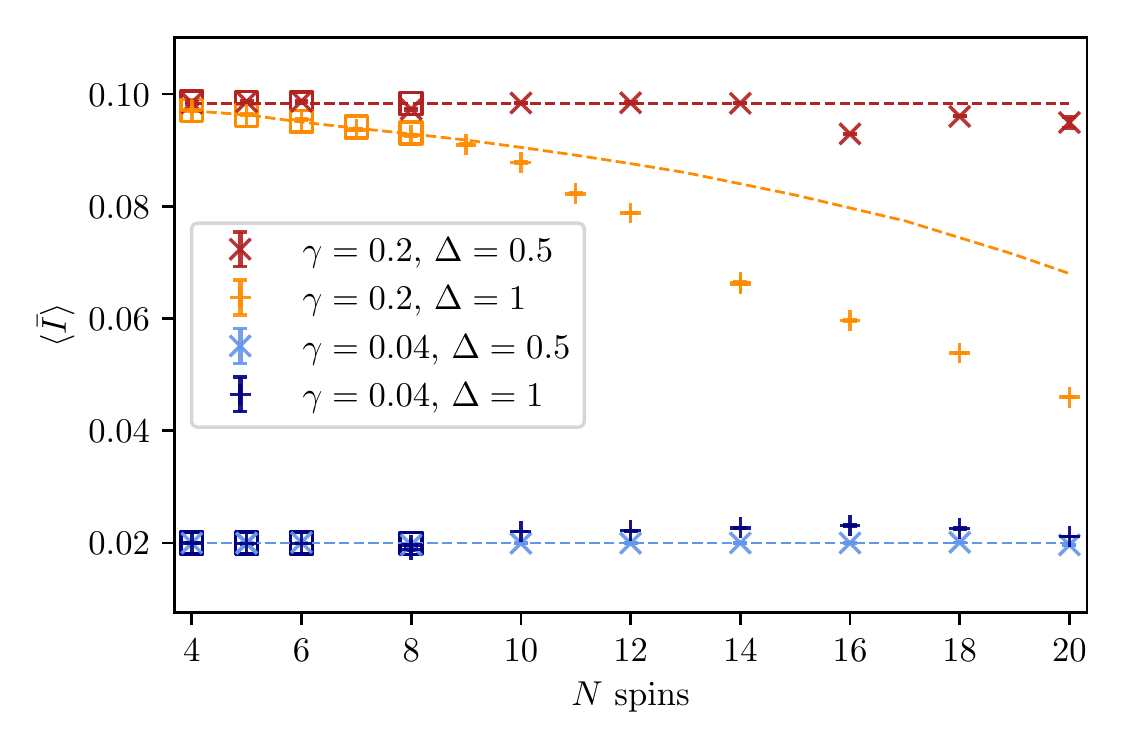}
	\caption{Mean spin current in a boundary driven Heisenberg chain (model B) as a function of chain length for different dissipation rates $\gamma$ and anisotropies $\Delta$ compared with exact diagonalization. The squares are results from exact diagonalization and the dashed lines resemble the current values from Ref.~\cite{prosen_exact_2011} including the length-independent analytical values for $\Delta = 0.5$.}
	\label{fig:current_B_anisoGamma_N} 
\end{figure}

For model B, described in section~\ref{sec:dissipative_driven_chain}, 
with only one Lindblad jump operator at each end of the chain, 
the  steady state current between neighbouring sites  
is independent of the position in the chain. That is not exactly true as long as the density matrix is not converged. We therefore compute the mean spin current $\bar{I}$, averaged over all nearest neighbour pairs in the chain. 
The result is displayed in Fig.~\ref{fig:current_B_anisoGamma_N} for different values of $\gamma$ and anisotropy $\Delta$.  
The anisotropic Heisenberg chain with $\Delta \le  0.5$ describes ballistic transport with 
a current that is independent of $N$ \cite{prosen_exact_2011} and hence is suitable to verify our numerical results for larger systems. 
For $\Delta=0.5$ we find very good agreement with the exact results also for systems up to 20 sites. 
For $\Delta=1$ we expect the same current values for small chains until spin diffusion
sets in and lets the current decrease with increasing $N$~\cite{prosen_exact_2011}. 
For small $\gamma$ this effect starts to appear at large chain lengths that are beyond our limited simulations. 
For $\gamma=0.2$ our simulation reproduces this effect, 
although at a decreased accuracy when compared to results from Ref.~\cite{prosen_exact_2011} displayed as the dashed lines. 
An explanation for this larger error at $\Delta=1$ is the following: If the steady state is not found exactly, contributions from the lowest eigenvectors of $\Li^\dagger \Li$ with non-zero eigenvalues remain. 
As $\Delta$ is increased, we noticed that these other eigenvectors have increasingly large contributions to the expectation value of the current, so that they have a greater influence on the spin current result of the steady-state approximation. In this particular system, this leads to a larger error, as can be seen in the $\Delta = 1$ result. 

\begin{figure}
	\centering
	\includegraphics[clip, trim=0cm 0cm 0cm 0cm,
	width=1\linewidth]{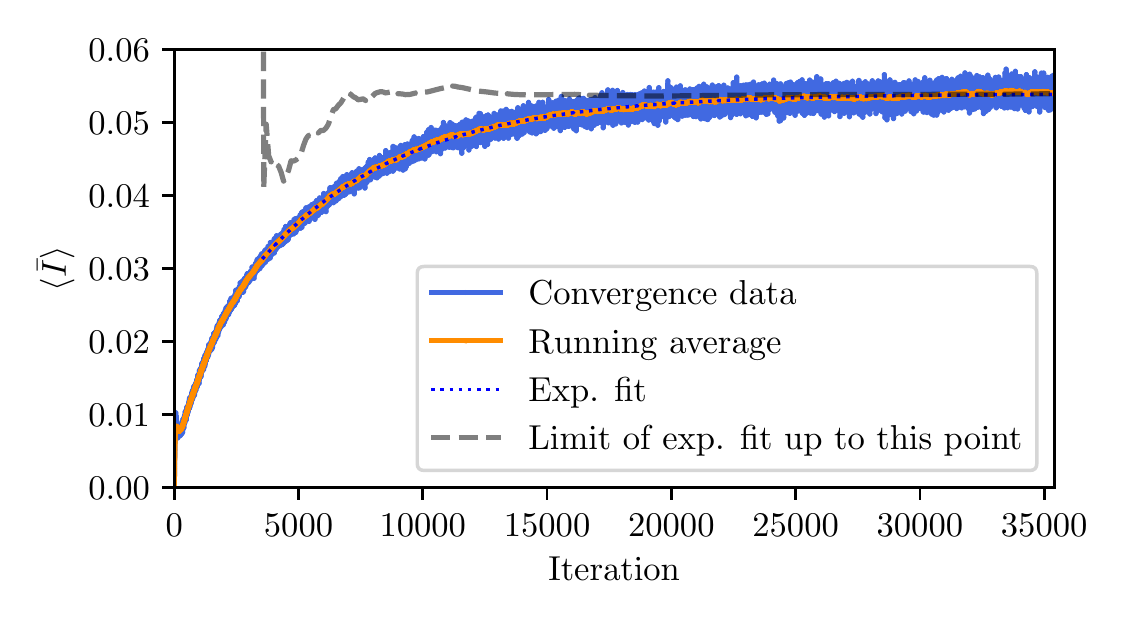}
	\includegraphics[clip, trim=-0.01cm 0cm -0.66cm 0cm,
	width=1\linewidth]{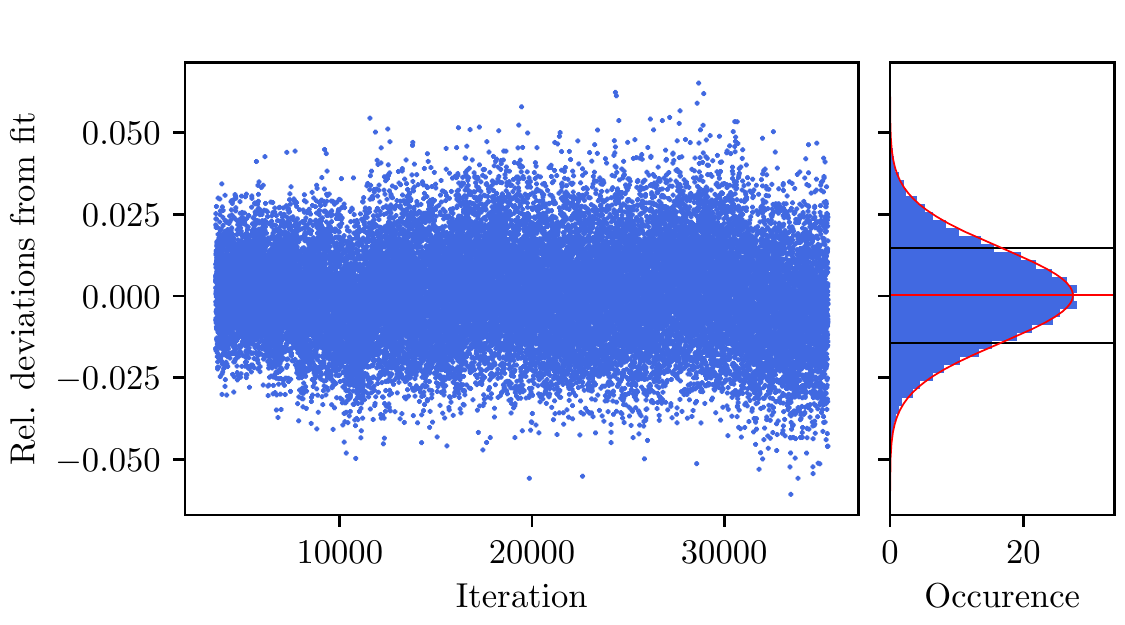}
	\caption{Top: 
A slower convergence of the spin current is observed for
 model B with $\gamma=0.2$, $\Delta=1$ (here: 18 spins).  
	A Linear convergence rate resulting in an exponential approach to a limit $I^*$ can been observed and verified by fitting the function $I(t) = I^* + C \lambda ^t$ to the data. The dashed line shows the limits $I^*$ obtained by only considering data up to this iteration. 
	Bottom: The deviations from the exponential fit are plotted and their distribution closely resembles a Gaussian (red curve). 
	}
	\label{fig:current_B_convergence} 
\end{figure}

For $\gamma=0.2$, $\Delta=1$ the system
also exhibited generally slower convergence compared to other
parameter values
we tested. Nevertheless,
after some iterations a linear convergence rate can be observed.
This leads to an exponential approach 
$I(t) = I^* + C \lambda ^t$ to a final value $I^*$ \cite{sidi_2003}.
In Fig.~\ref{fig:current_B_convergence} we illustrate such a convergence curve together with an exponential fit and its Gaussian fluctuations.  
The convergence data fluctuates about a mean value due to the Monte-Carlo sampling and possibly due to updates in the parameters. The deviations from the mean nicely follow a Gaussian distribution as expected for MC sampling. 
The extrapolated results $I^*$ approach the final value 
already at an early stage of the
optimization. This could therefore 
potentially be used as a stopping condition. 
The converged value is independent of the 
initial parameters, as long as they are not chosen too large so that the procedure remains numerically stable. 
Also an increase in the number of parameters of the neural network, achieved by doubling $\alpha$ and $\beta$, leads to the same results. 
In all these cases, the SR optimization method produces similar values for the spin current but takes longer to converge.
The discrepancy with the exact results observed in Fig.~\ref{fig:current_B_anisoGamma_N} 
 suggests that the expressibility of this RBM network ansatz may be not sufficient for this setup.

For these  
parameter values we used $\alpha=\beta=4$ for the hidden and ancillary node densities  
 whereas in all other simulations $\alpha=\beta=2$ was used. 
The sample sizes were in the range from $2000$ for 8 spins to $8000$ for 18 spins. 
The number of diagonal samples in Eq.~\eqref{eq:observable} was increased to $10^5$
for the final evaluation of the observables. 

\section{Conclusion}

We have investigated the applicability of neural density operators based on RBM as approximation of the non-equilibrium steady-state  of asymmetrically driven spin chains. 
We have focused primarily on magnetization and spin current, and have shown that the magnetization alone is not a good indicator of the quality of the approximation. 

Using more advanced optimization techniques we have shown that high fidelity approximations are achievable. 
We have also addressed the previously reported inadequacy of Metropolis Monte-Carlo sampling in such systems and propose a new sampling method to overcome these problems  and eventually obtained good results for the NESS spin current for some larger model systems. 
For some values of the parameters, accurate results for the spin current are harder to achieve due to a slower convergence and a larger contribution from slowly decaying density matrices.
In some situations the spin current consistently converges to an inaccurate value. This
 may suggest a lack of expressibility of the RBM ansatz function in these cases.

The proposed optimization method provides orders of magnitude faster convergence for most of the tested asymmetric systems and, in some cases, enables a viable solution in the first place. Hereby, neural network quantum states can be extended to quantum transport in asymmetrically driven open spin chains.

\section*{Acknowledgments}
The implementation was based on the Jax \cite{jax2018github} and NetKet \cite{carleo_machine_2019} libraries. 
This research was funded in part by the Austrian Science Fund (FWF): P 33165-N, and by NaWi Graz. 

\newpage
\onecolumngrid
\section*{Appendix A: The RBM Mixed Density Matrix Ansatz}

The final expression we used as the RBM density matrix ansatz in Eq.~\eqref{eq:rho_ansatz} (see Ref.~\cite{torlai_latent_2018}) can be written as
\begin{align}
\rho_\ve{\lambda\mu}(\si,\si')  
	&= \exp \left({\Gamma_\ve{\lambda}^{(+)}(\si,\si') + i \Gamma_\ve{\mu}^{(-)}(\si,\si') + \Pi_\ve{\lambda\mu}(\si,\si') } 
	\right) \\
\Gamma_X^{(\pm)}(\si, \si')
	&= \frac{1}{2} 	\left( \sum_j \log  G(y_j^X(\si)) 
	\pm \log  G(y_j^X(\si')) 
	+ \sum_i b_i^X (\sigma_i \pm \sigma'_i) 
	\right) \\
\Pi_{\ve{\lambda\mu}}(\si, \si')
	&= \sum_k \log \left[
	G \left(
	\frac{1}{2} \sum_l U^\lambda_{lk} (\sigma_l + \sigma'_l) + 
	\frac{i}{2}  \sum_l U^\mu_{lk} (\sigma_l - \sigma'_l) + 
	d^\lambda_k
	\right)
	\right]
\end{align}
where $y^X_j(\si) = \sum_i W^X_{ij} \sigma_i + c^X_j$ and  $G(x) = 2\cosh(x)$ as a nonlinear activation function when considering $\sigma \in \{-1, 1\}$.

\twocolumngrid
\bibliographystyle{ieeetr}
\bibliography{NeuralQuantumNetwork,additionalReferences}

\end{document}